\documentclass[showpacs,aps,floatfix,epsfig]{revtex4}
\usepackage{graphicx}
\usepackage{amsmath}
\usepackage{amssymb}
\usepackage{bm}
\begin{document}
\title{Susceptibilities with multi-quark interactions in PNJL model} 
\author{Abhijit Bhattacharyya}
\email{abphy@caluniv.ac.in}
\author{Paramita Deb}
\email{paramita.deb83@gmail.com}
\affiliation{Department of Physics, University of Calcutta,
92, A. P. C. Road, Kolkata - 700009, INDIA}
\author{Anirban Lahiri}
\email{anirbanlahiri.boseinst@gmail.com}
\author{Rajarshi Ray}
\email{rajarshi@bosemain.boseinst.ac.in}
\affiliation{Center for Astroparticle Physics \&
Space Science, Bose Institute, Block-EN, Sector-V, Salt Lake, Kolkata-700091, INDIA 
 \\ \& \\ 
Department of Physics, Bose Institute, \\
93/1, A. P. C Road, Kolkata - 700009, INDIA}

\begin {abstract}
We have investigated the fluctuations and the higher order susceptibilities
of quark number, isospin number, electric charge and strangeness at 
vanishing chemical potential for 2+1 flavor Polyakov loop extended
Nambu--Jona-Lasinio model. The calculations are performed for 
the bound effective potential in the quark sector requiring up to eight
quark interaction terms. These have been contrasted to the lattice 
results which currently have somewhat heavier quarks in the light flavor
sector. The results show sufficient qualitative agreement. For comparison
we also present the results obtained with the conventional 
effective potential containing upto six quark interaction terms.
\end{abstract}
\maketitle

{\section{Introduction}}
 Confinement and chiral symmetry breaking are the most fundamental properties 
of strong interaction physics at low temperature and density, where the physics
is mainly governed by the non-perturbative QCD. In principle, the  
deconfinement phase transition and chiral phase transition are defined in 
two extreme limits of current quark mass. Deconfinement phase transition and
its order parameter is well defined for infinite current quark mass and chiral
phase transition is exact for zero quark mass. But in real world with a finite 
value of quark masses, the nature of these two phase transitions is an open
question. Another important feature of the QCD phase diagram is the existence
of the critical end point (CEP), where first order phase transition, from
hadronic phase to quark-gluon-plasma (QGP) phase, ends. However, the exact
location of CEP is still unknown. Investigation 
of these properties of strongly interacting matter are necessary to understand
the various astrophysical and cosmological scenario.

  The experimental explorations to understand the properties of strongly 
interacting matter has been studied at Relativistic Heavy-ion-collider (RHIC),
Brookhaven \cite{adcox}  
where the heavy nuclei collide with each other at relativistic energies 
to form hot and dense strongly interacting matter. More data are 
expected from LHC and  
FAIR in future. But to analyze the data from the experiment, we need
a thorough understanding of the theory of strong interaction physics.

  Due to our limited knowledge of the non-perturbative physics, QCD, which 
is the theory of strong interaction, can not be used to study the phase 
transition picture. In this regard Lattice QCD (LQCD) provides the 
most direct approach to study QCD at high temperature  \cite{boyd,engels,
fodor1,fodor2,allton1,
allton2,allton3,forcrand, aoki1,aoki2,arriola1,arriola2,arriola3,arriola4}. However LQCD
has its own restrictions due to the discretization of space-time. Furthermore,
at finite chemical potential, LQCD faces the well known sign problem. 

  Another approach to study low energy limit of QCD and the QCD phase 
transition is to use effective theories of QCD.  
Polyakov loop extended Nambu--Jona-Lasinio model (PNJL) is one of the 
successful approach which combines the confinement and chiral symmetry 
breaking properties in a simple formalism.
There have been series of work to study the thermodynamic properties
of strongly interacting matter using PNJL model for both 2 flavor and
2+1 flavor \cite{fuku1,ratti1,pisarski,fuku2,
ratti2,gatto,ghosh,rayvdm,deb1}. These studies suggest that this model
reproduces the zero density lattice data quite successfully.
However, the vacuum of the NJL part of the PNJL model seems to to be
unbound in a 2+1 flavor scenario. A plausible solution of this problem
has been proposed by Osipov {\it{et.al}} using eight-quark interaction
term \cite{osipov1,osipov2,osipov3,osipov4}.
Also the 2 flavor PNJL model have been studied in Ref. 
\cite{kashiwa1,kashiwa2} with eight-quark interaction. In our previous work, we developed
2+1 flavor PNJL model with eight-quark interaction
terms in the Lagrangian with three-momentum cutoff regularisation \cite{deb2}.
\par
The thermodynamic aspect of the phase transition from hadronic phase to QGP
phase can be understood properly if we study the thermodynamic variables like
quark number susceptibility (QNS), isospin number susceptibility (INS), 
specific heat $(C_V)$ and speed of sound ($v_s$) etc. Susceptibilities are 
related to fluctuations via the fluctuation-dissipation theorem.
 A measure of the intrinsic
statistical fluctuations in a system close to thermal equilibrium is provided
by the corresponding susceptibilities. At zero chemical potential, charge 
fluctuations are sensitive indicators of the transition from hadronic matter
to QGP. Also the existence of the CEP can be 
signalled by the divergent fluctuations.
For the small net baryon number, which
can be met at different experiments, the transition from hadronic to QGP phase
is continuous and 
the fluctuations are not expected to lead any singular behavior.  
Recently, the computations on the lattice have been performed for
many of these susceptibilities at zero chemical potentials 
\cite{gottlieb,gavai,bernard1,bernard2}. It was shown that
at vanishing chemical potential the susceptibilities
rise rapidly around the continuous crossover transition region.

 The study of higher order moments of fluctuations are also necessary to 
locate the transition point more accurately. In 2 flavor QCD, it has been 
shown that the quark number and isospin number fluctuations increase with 
temperature and their fourth moments start to show pronounced peaks in the 
transition region from low to high temperature \cite{allton3,ejiri}. 
In fact the higher order coefficients become increasingly sensitive in the 
vicinity of phase transition.
Fluctuations are computed with respect to
the quark chemical potential in 2 flavor PNJL model with three-momentum cutoff
regularisation \cite{ghosh,roessner1,friman,ray2}.
Also QNS at finite density has been estimated in some works within 2 flavor 
PNJL model \cite{roessner2}. Recently the idea of the Taylor
expansion in terms of chemical potential for PNJL model has been computed within
the constraint that the net strange quark density is zero, which is the case in
ultra relativistic heavy ion collision \cite {fuku3}.
There have been recent calculations towards the 
fluctuations in 2+1 flavor Lattice QCD \cite{cheng1,cheng,fodor10} 
and also in the 2+1 flavor PNJL model taking upto six-quark determinant
interaction terms \cite{fuku2,wu1}.
Similar calculations have been carried out in Polyakov loop coupled quark-meson
(PQM) model \cite{schaefer1,schaefer2,schaefer3,schaefer4} and its renormalization
group improved version \cite{skokov}.

In this paper we have investigated the susceptibilities within 2+1 flavor PNJL model
framework with two different kinds of NJL interaction. In one case we
take the conventional form which takes into account upto six-quark
determinant interaction, which we will call as model A. In the other case an
extra eight-quark interaction term is added to make the effective potential
bound, which will be hereafter denoted by model B.  
We have also studied the specific heat and the speed of sound.
Specific heat is related to the event-by-event temperature fluctuations
\cite{stod} and mean transverse momentum fluctuations \cite{korus} in 
heavy-ion reactions.
These fluctuations show diverging behavior near the critical end point (CEP).
The speed of sound determines the flow properties in heavy-ion reactions 
\cite{sorge, kolb}.

 Our paper is organized as follows: In Sec. II, the basic formalism of the PNJL
model and also the calculation of fluctuations have been discussed. 
Various thermodynamic quantities such as specific heat, speed of  
sound are also defined in this section. 
In the next section we describe our results and compared with the recent lattice
data and also with other models like Polyakov extended quark-meson model (PQM).
In the last section we conclude.
 
\vskip 0.3in
\section {Formalism}
{\subsection {Thermodynamic Potential}}

The PNJL model was formulated to study the chiral properties and the
confinement physics of the QCD phase transition at finite
temperature and density. In this model quark dynamics is studied with a background
gauge field having only the temporal component. For a detailed review of the 
PNJL model with 2 flavor and 2+1 flavor see Ref.
\cite{fuku1,ghosh,gatto,roessner1,deb1,friman,roessner2,skokov,ray2,fuku2,fuku3,wu1}.
Osipov {\it{et.al}} pointed out that the effective potential of the NJL part is seems to be unbound in the conventional form and they have introduced the eight-quark interaction terms
in NJL model to stabilize the vacuum \cite{osipov1,osipov2,osipov3,osipov4}. 
Also the 2 flavor PNJL model have been studied in Ref. 
\cite{kashiwa1,kashiwa2} with eight-quark interaction.
In our previous paper we developed the 2+1 PNJL model with eight-quark 
interactions within three-momentum cutoff scheme \cite{deb2} for the first 
time. We have reproduced the 
thermodynamic properties of QCD, calculated on Lattice, at zero baryon density 
quite satisfactorily.  The thermodynamic
potential for the multi-fermion interaction in the mean field 
approximation (MFA) of the PNJL model can be written as \cite{deb2}, 
\begin {align}
 \Omega &= {\cal {U^\prime}}[\Phi,\bar \Phi,T]+2{g_S}{\sum_{f=u,d,s}}
            {\sigma_f^2}-{{g_D} \over 2}{\sigma_u}
          {\sigma_d}{\sigma_s}+3{{g_1}\over 2}({\sum_{f=u,d,s}}{\sigma_f}^2)^2\nonumber\\
           &+3{g_2}{\sum_{f=u,d,s}}{\sigma_f^4}-6{\sum_{f=u,d,s}}{\int_{0}^{\Lambda}}
     {{d^3p}\over{(2\pi)}^3} E_{f}\Theta {(\Lambda-{ |\vec p|})}\nonumber \\
       &-2T{\sum_{f=u,d,s}}{\int_0^\infty}{{d^3p}\over{(2\pi)}^3}
       \ln\left[1+3(\Phi+{\bar \Phi}e^{-{(E_{f}-\mu_f)\over T}})
       e^{-{(E_{f}-\mu_f)\over T}}+e^{-3{(E_{f}-\mu_f)\over T}}\right]\nonumber\\
       &-2T{\sum_{f=u,d,s}}{\int_0^\infty}{{d^3p}\over{(2\pi)}^3}
        \ln\left[1+3({\bar \Phi}+{ \Phi}e^{-{(E_{f}+\mu_f)\over T}})
       e^{-{(E_{f}+\mu_f)\over T}}+e^{-3{(E_{f}+\mu_f)\over T}}\right]
\end {align}
$g_S$ and $g_D$ are the four-quark and six-quark coupling constants respectively 
and $g_1$ and $g_2$ are the eight-quark coupling constants.
Here $\sigma_f=\langle{\bar \psi_f} \psi_f\rangle$
denotes chiral condensate of the quark with flavor $f$ and
$E_{f}=\sqrt {p^2+M^2_f}$ is the single quasi-particle energy.
Here, constituent mass $M_f$ of flavor $f$ is given by the self-consistent
gap equation;
\begin{equation*}
M_f=m_f-2g_S\sigma_f+{{g_D}\over 2}\sigma_{f+1}\sigma_{f+2}-2g_1\sigma_f
(\sigma_u^2+\sigma_d^2+\sigma_s^2)-4g_2\sigma_f^3
\end{equation*}
where $f$, $f+1$ and $f+2$ take the labels of flavor $u$, $d$ and $s$
in cyclic order. So, when $f=u$ then $f+1=d$ and $f+2=s$ and so on.
In the above integrals, the vacuum 
integral has a cutoff $\Lambda$ whereas the medium dependent integrals have
been extended to infinity.
Polyakov loop being the normalized trace of the Wilson line $\textbf{L}$,
which is an SU(3) matrix, should lie in the range 0$\leqslant\Phi\leqslant$1.
But earlier studies of PNJL model \cite{ghosh,ray2,deb1} show that the 
$\Phi$ becomes greater than 1 above 2$T_C$. To solve this problem one has to
take a proper Jacobian of transformation from the matrix valued field 
$\textbf{L}$ to the complex valued field $\Phi$. This will constrain 
the value of $\Phi$
within 1. Thus one has to modify the Polyakov loop potential by introducing
Vandermonde (VdM) term.
The necessity of the modification was reflected through an excellent 
agreement of flavour mixing effects in PNJL model with lattice data as shown in Ref. \cite{rayvdm}.  
There are a few extra terms in ${\cal U}(\Phi,\bar\Phi)$ in this prescription
as compared to other recent works \cite{fuku2,roessner1}.
These extra terms are put on the basis of the global Z(3)
symmetry of the Polyakov loop potential and therefore their
presence seem to be quite natural though may not be absolutely necessary
as shown by  \cite{fuku2,roessner1}. 
The modified  potential $\cal {U^\prime}$ can be expressed as ,
\begin{equation}
{{\cal {U^\prime}}(\Phi,\bar \Phi,T)\over {T^4}}=
 {{\cal U}(\Phi,\bar \Phi,T)\over {T^4}}-\kappa \ln[J(\Phi,{\bar \Phi})]
\label {uprime}
\end{equation}
where ${\cal U}(\Phi,\bar \Phi,T)$ is the Landau-Ginzburg type potential 
given by \cite{ratti1},
\begin{equation}
  {{\cal U}(\Phi, \bar \Phi, T) \over {T^4}}=-{{b_2}(T) \over 2}
                 {\bar \Phi}\Phi-{b_3 \over 6}(\Phi^3 + \bar \Phi^3)
                 +{b_4 \over 4}{(\bar\Phi \Phi)}^2
\label{LG_pot}
\end{equation}
with,
\begin{equation}
     {b_2}(T)=a_0+{a_1}({{T_0}\over T})+{a_2}({{T_0}\over T})^2+
              {a_3}({{T_0}\over T})^3,
\label{b2}
\end{equation}
 and $b_3$, $b_4$ being constants. 
$J(\Phi, {\bar \Phi})$ in eqn. (\ref{uprime}) is known as 
VdM determinant \cite{rayvdm}, is given by,
\begin {equation*}
J[\Phi, {\bar \Phi}]=(27/24{\pi^2})(1-6\Phi {\bar \Phi}+\nonumber\\
4(\Phi^3+{\bar \Phi}^3)-3{(\Phi {\bar \Phi})}^2)
\end{equation*}
$\kappa$ is a phenomenological constant.
Polyakov loop $\Phi$ and its charge conjugate $\bar \Phi$ is defined as, 
\begin {equation*}
\Phi = (\rm{Tr}_c L)/N_c, {\hspace{0.3in}} {\bar \Phi} = (\rm{Tr}_c L^\dagger)/N_c
\end {equation*}

The parameter $T_0$ is taken as $190~ \rm MeV$, whereas the lattice determines 
its value to be $270~ \rm MeV$ for pure gauge theory. The reason to
take a lower value of $T_0$ is to get the crossover temperature ($T_c$) 
consistent with the lattice data.
In this work we have taken the parameter set obtained in our previous paper
\cite{deb2}. For fixing the parameters $m_s,\Lambda,g_S,g_D,g_1,g_2$
we have used the following physical conditions;
\begin{equation*}
m_\pi=138 ~ {\rm MeV} ~~~ f_\pi=93 ~ {\rm MeV} ~~~ m_K=494 ~ {\rm MeV} ~~~ \\
f_K=117 ~ {\rm MeV} ~~~ m_\eta=480 ~ {\rm MeV} ~~~ m_{\eta\prime}=957 ~ {\rm MeV}
\end{equation*}
and $m_u$ is kept fixed at 5.5 MeV.
The parameters are given in table \ref{table1} for both PNJL model A and PNJL model B.
\begin{table}
\begin{center}
\begin{tabular}{|c|c|c|c|c|c|c|c|c|c|c|}
\hline
Interaction &$ m_u $&$ m_s $&$ \Lambda $&$ g_S \Lambda^2 $&$ g_D \Lambda^5 $&$
g_1 \times 10^{-21} $&$ g_2 \times 10^{-22}$&$ \kappa $&$
T_C $ \\
 &$\rm (MeV)$& $\rm (MeV)$&$\rm (MeV)$&  &  &$ \rm (MeV^{-8})$ & $ \rm (MeV^{-8})$ &
  & $\rm (MeV)$ \\
\hline
                                                                                
$\rm {Model A} $&$ 5.5 $&$ 134.758 $&$ 631.357 $&$ 3.664 $&$ 74.636
$&$ 0.0 $&$ 0.0 $&$ 0.13 $&$ 181 $\\

$\rm {Model B}$ & $ 5.5 $&$ 183.468 $&$ 637.720 $&$ 2.914 $&$ 75.968
$&$ 2.193 $&$ -5.890 $&$ 0.06 $&$ 169 $\\
 
\hline
\end{tabular}
\caption{ Parameters and $T_C$ for model A and model B Lagrangians.}
\label{table1}
\end{center}
                                                                                
\end{table}

\par
For the Polyakov loop potential we choose the
parameters which reproduces the lattice data of pure
gauge thermodynamics \cite{Boyd}. It was shown in Ref. \cite{ratti1}
that, pure gauge Lattice QCD data of  scaled pressure, entropy and 
energy density are reproduced extremely well in Polyakov loop model using 
the ansatz (\ref{LG_pot}) and (\ref{b2}) with parameters summarized below,
\begin {align}
       a_0=6.75,    a_1=-1.95,  a_2=2.625,
      a_3=-7.44,   b_3=0.75,  b_4=7.5 \nonumber
\end {align}

\vskip 0.2in
{\subsection{Taylor expansion of pressure}}
   The pressure of the strongly interacting matter can be written as,
\begin {equation}
P(T,\mu_q,\mu_Q,\mu_S)=-\Omega (T,\mu_q,\mu_Q,\mu_S),
\label{pres}
\end {equation}
where $T$ is the temperature, $\mu_q$ is the quark chemical potential, 
$\mu_Q$ is the charge chemical potential and $\mu_S$ is the strangeness chemical potential. 
From the usual thermodynamic
relations we can show that the first derivative of pressure with respect to
$\mu_q$ gives the quark number density and the second derivative is the 
quark number susceptibility (QNS). 

 Our first job is to minimize the thermodynamic potential numerically with
respect to the fields $\sigma_u$, $\sigma_d$, $\sigma_s$, $\Phi$ and 
$\bar \Phi$. The values of the fields can then be used to evaluate the 
pressure using the equation (\ref{pres}).
Then we can expand the scaled pressure at a given temperature in a Taylor
series for the chemical potentials $\mu_q$, $\mu_Q$, $\mu_S$ as,
\begin {equation}
{p(T,\mu_q,\mu_Q,\mu_S)\over{T^4}}=\sum_{i,j,k} c_{i,j,k}^{q,Q,S} 
  ({\mu_q\over T})^i ({\mu_Q\over T})^j ({\mu_S\over T})^k
\label{pres_Taylor}
\end{equation}
where,
\begin {equation}
c_{i,j,k}^{q,Q,S}(T)={{1\over {i! j! k!}} {\partial^i \over 
\partial ({\mu_q\over T})^i} {\partial^j \over 
\partial({\mu_Q\over T})^j} {\partial^k {(P/T^4)} \over 
\partial({\mu_S\over T})^k}}\Big|_{\mu_{q,Q,S}=0}
\end{equation}
The flavor chemical potentials  $\mu_u$, $\mu_d$, $\mu_s$ are related to 
$\mu_q$, $\mu_Q$, $\mu_S$ by,
\begin {equation}
  \mu_u=\mu_q+\frac{2}{3}\mu_Q,~~~ 
  \mu_d=\mu_q-\frac{1}{3}\mu_Q,~~~
  \mu_s=\mu_q-\frac{1}{3}\mu_Q-\mu_S
\label{mureln1}
\end {equation}
It should be mentioned that one can also choose the independent chemical potentials
as $\mu_q$, $\mu_I$, $\mu_S$. Then eqn.(\ref{mureln1}) becomes,
\begin {equation}
  \mu_u=\mu_q+\mu_I,~~~ 
  \mu_d=\mu_q-\mu_I,~~~
  \mu_s=\mu_q-\mu_S
\label{mureln2}
\end {equation}
where, $\mu_I$ is the isospin chemical potential.
Eq.(\ref{pres_Taylor}) is a general expression of Taylor expansion of pressure for different
chemical potentials. Since in this paper we are only concerned with the 
diagonal terms of the expansion, we can write the above equation 
in simpler way as,
\begin {equation}
{p(T,\mu_X)\over{T^4}}=\sum^{\infty}_{n{=0}}c_{n}^{X}(T) 
                      ({\mu_X\over T})^n
\end{equation}
where,
\begin {equation}
c_{n}^{X}(T)={{1\over {n!}} {\partial^n {(P(T,\mu_X)/T^4)} 
\over \partial ({\mu_X\over T})^n}}\Big|_{\mu_X=0}
\end{equation}
Where X is $q$, $Q$ or $I$ and $S$.
Here we will use the expansion around $\mu_X=0$, where the odd terms
vanish due to CP symmetry. In this work we evaluate the expansion coefficients
up to eighth order. To obtain the Taylor coefficients, first the pressure 
is obtained as a function of $\mu_X$ for each value of 
T, then fitted to a polynomial about $\mu_X=0$. All orders of derivatives 
are then obtained from the coefficients of the polynomial extracted from 
the fit. For the stability of the fit we have checked the values of 
least squares.
\vskip 0.2in
{\subsection {Specific heat and speed of sound}}
 We have studied the specific heat $C_V$, which is important to find the 
location of CEP, the speed of sound, which determines the flow properties
in heavy-ion reactions. The energy density $\epsilon$ is obtained from
the thermodynamic potential $\Omega$ as,
\begin{equation}
\epsilon={-T^2 {\partial {(\Omega/T)} \over {\partial T}}}\Big|_V\nonumber\\
        ={-T {\partial \Omega \over \partial T}}\Big|_V+ \Omega
\end{equation}
The specific heat is defined as the rate of change of energy density with 
temperature at constant volume, which is given by,
\begin{equation}
C_V={\partial \epsilon \over \partial T}\Big|_V \nonumber\\
    ={-T {{\partial^2 \Omega}\over \partial T^2}}\Big|_V.
\end{equation}
For a continuous phase transition near CEP, it is expected that $C_V$
shows a diverging behavior, which will translate into highly enhanced 
transverse momentum fluctuations or highly suppressed temperature fluctuations.
The square of speed of sound at constant entropy $S$ is given by,
\begin{equation}
v_s^2 = \left . {\partial P \over \partial \epsilon} \right |_S 
      = \left . {\partial P \over \partial T} \right |_V \left /
        \left . {\partial \epsilon \over \partial T} \right |_V \right .
      = \left . {\partial \Omega \over \partial T} \right |_V \left /
        \left . T {\partial^2 \Omega \over \partial T^2} \right |_V 
        \right .  ~~~.
\end{equation}
Divergence in specific heat near CEP means vanishing speed of sound.

\vskip 0.3in
{\section{Result}}
We now present the coefficients of the Taylor expansion of pressure for 2+1
flavor PNJL model with model A and model B Lagrangians and make 
a comparative study between the quark number susceptibility (QNS), 
isospin number susceptibility (INS), charge and strangeness susceptibility
and their higher order derivatives.
We then compare the specific heat and the speed of sound for both 
Lagrangians. We also compare our results with the recent
lattice data available for 2+1 flavor with $N_\tau=6$ \cite{cheng} and also
with that of Polyakov extended quark-meson model (PQM) 
\cite{schaefer1,schaefer2,schaefer3,schaefer4}.
\vskip 0.2in
{\subsection{Coefficients of the Taylor expansion}}
\begin{figure}[t]
\centering
\includegraphics[scale=0.9]{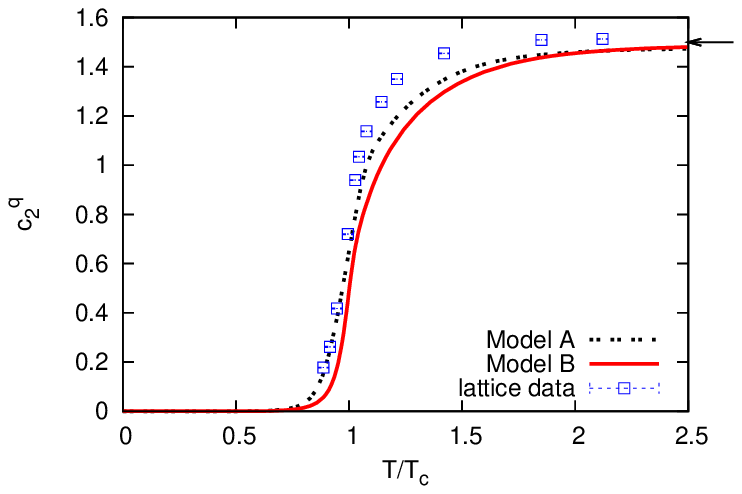}
\includegraphics[scale=0.9]{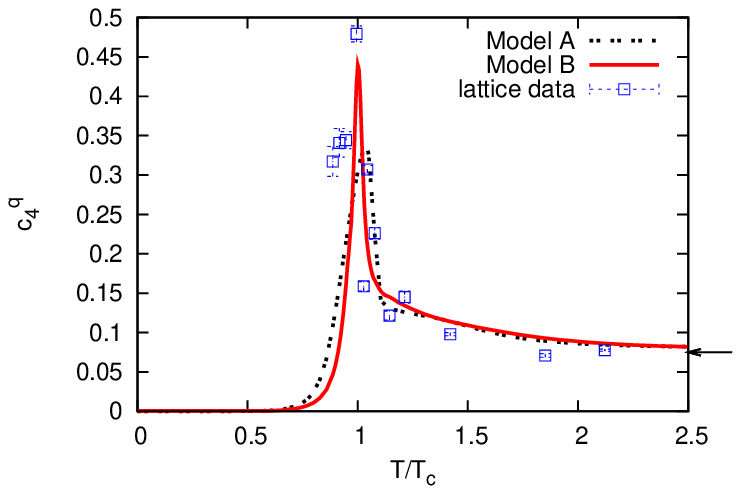}
\includegraphics[scale=0.9]{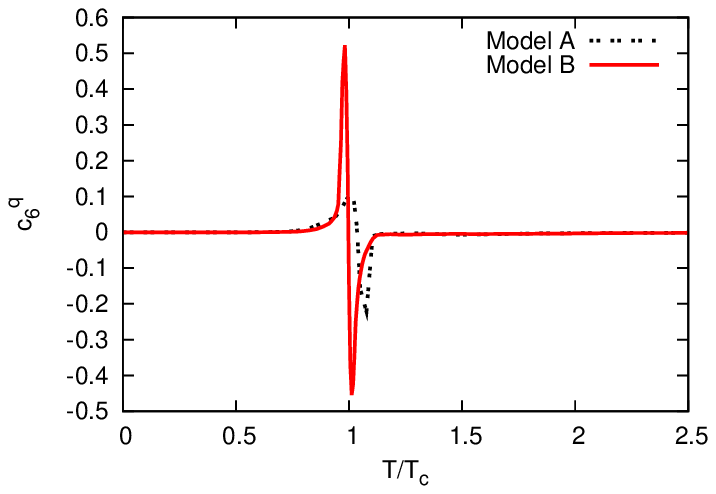}
\includegraphics[scale=0.9]{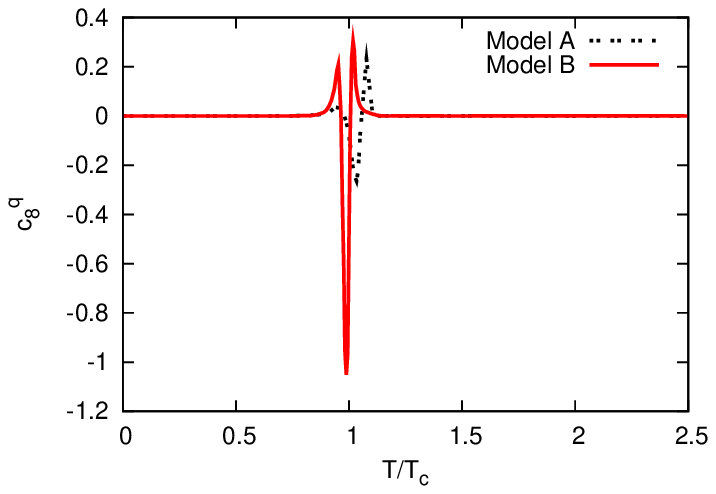}
\caption{(color online). Variation of $c_2$, $c_4$, $c_6$ and $c_8$ 
with $T/T_C$, for 
$\mu_X=\mu_q$ for models A and B. The arrows on the
right show the corresponding SB limits. The lattice are data taken
from Ref. \cite{cheng}.}
\label {qns}
\end{figure}

\begin{figure}[t]
\centering
\includegraphics[scale=0.9]{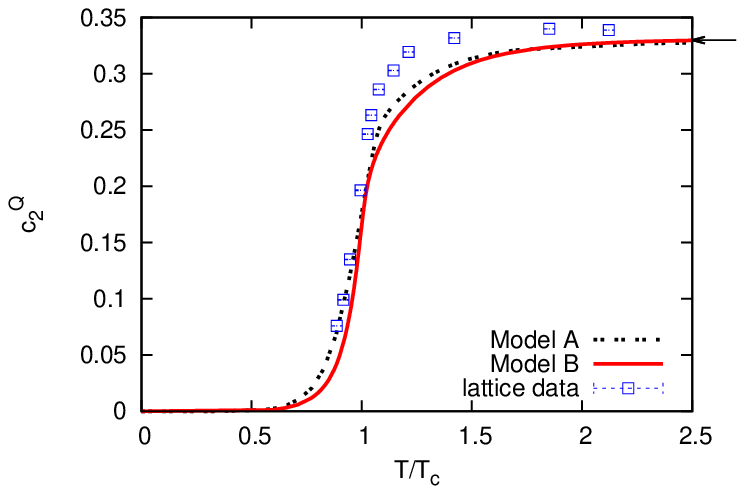}
\includegraphics[scale=0.9]{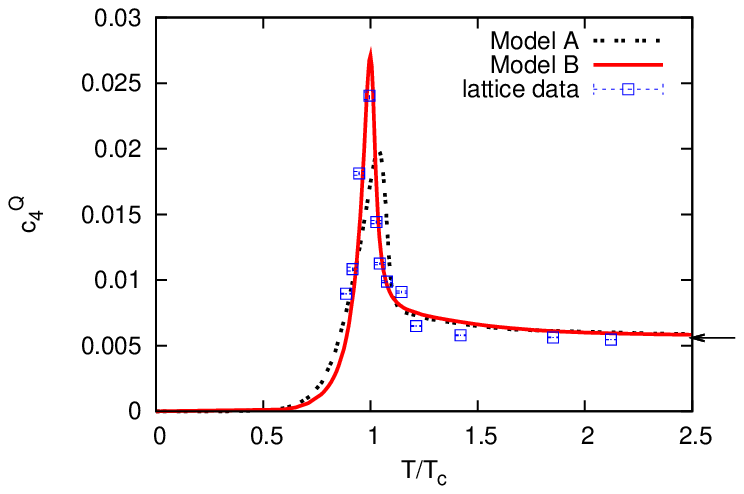}
\includegraphics[scale=0.9]{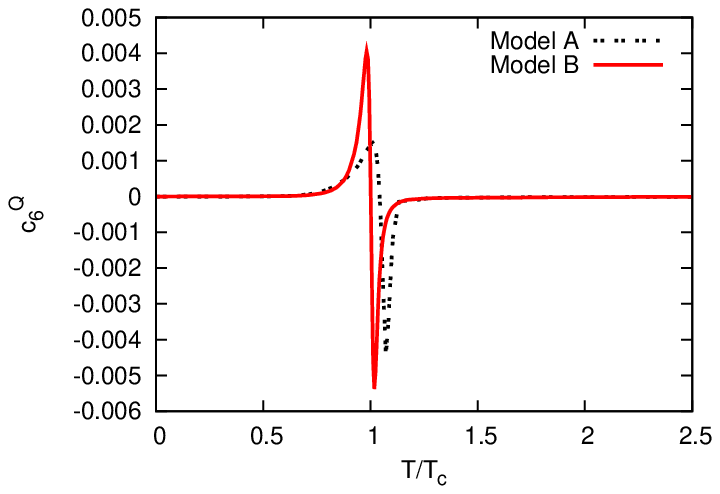}
\includegraphics[scale=0.9]{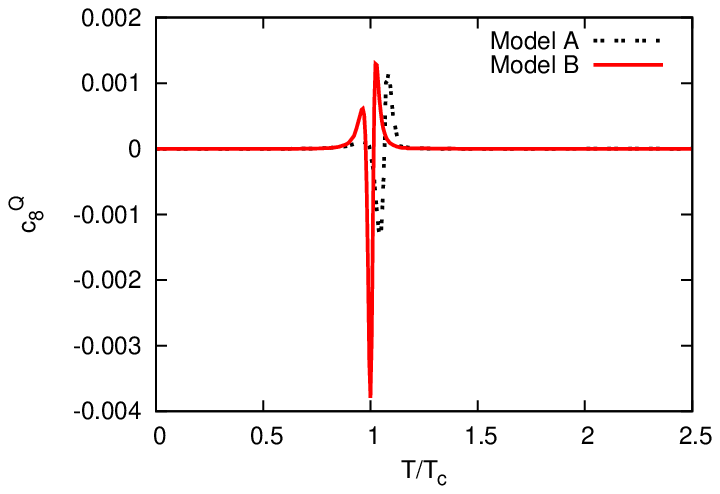}
\caption{(color online). Variation of $c_2$, $c_4$, $c_6$ and  $c_8$ 
with $T/T_C$, for
$\mu_X=\mu_Q$  for models A and B. The arrows on the
right show the corresponding SB limits. The lattice data are taken
from Ref. \cite{cheng}.}
\label {charge}
\end{figure}
\begin{figure}[t]
\centering
\includegraphics[scale=0.9]{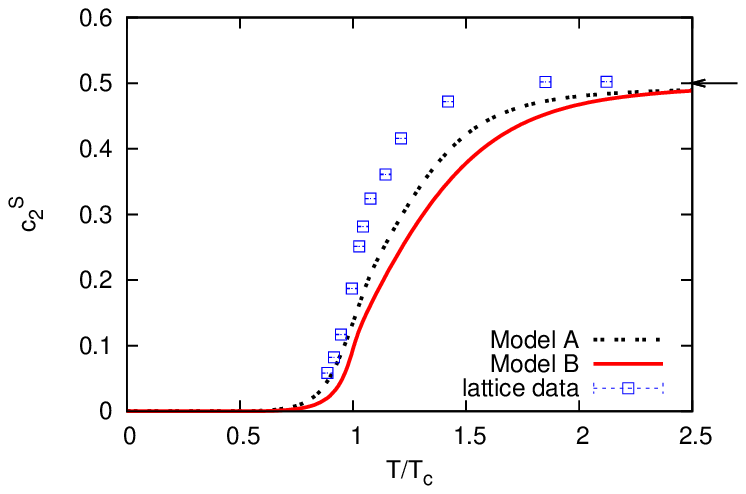}
\includegraphics[scale=0.9]{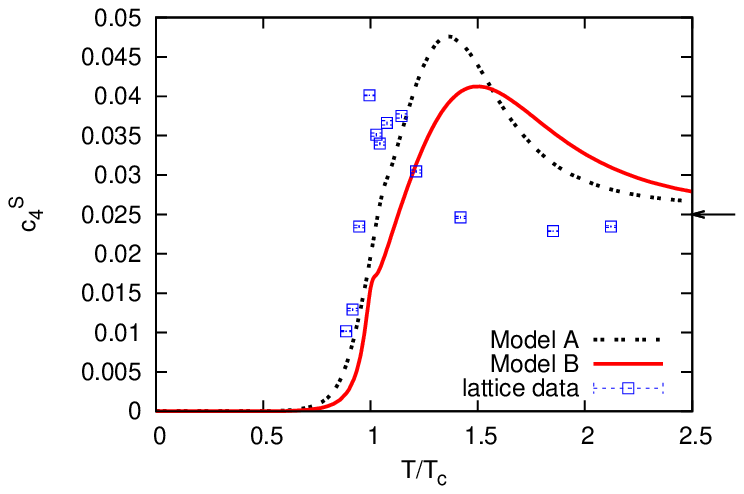}
\includegraphics[scale=0.9]{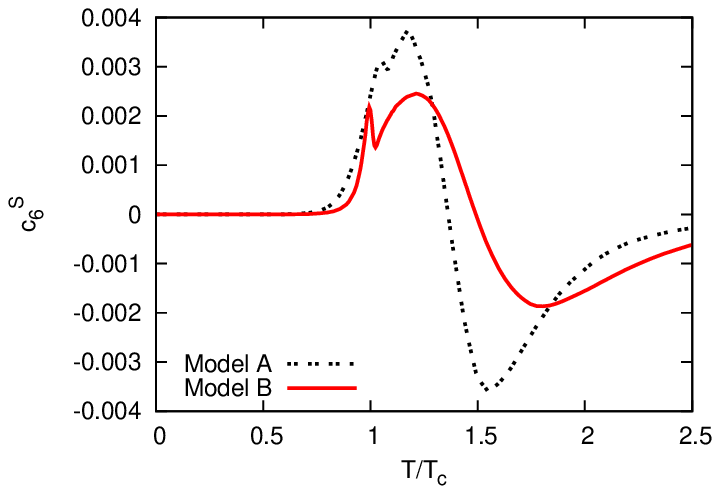}
\includegraphics[scale=0.9]{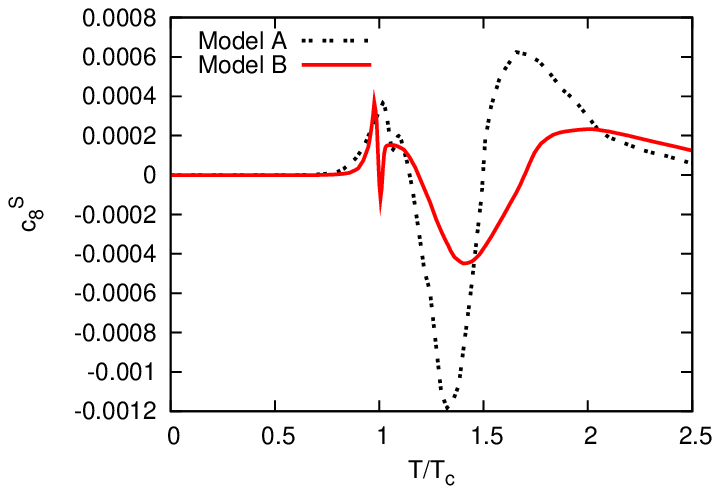}
\caption{(color online). Variation of $c_2$, $c_4$, $c_6$ and $c_8$ 
with $T/T_C$, for
$\mu_X=\mu_S$  for models A and B. The arrows on the
right show the corresponding SB limits. The lattice are data taken
from Ref. \cite{cheng}.}
\label {sns}
\end{figure}
The pressure is fitted to a 
polynomial in $\mu_X$ using the \textquotedblleft  gnuplot\textquotedblright
\cite{gnu}
program at different values of temperature. 
Here we consider to take maximum eighth order term in the 
polynomial in $\mu_X$. We restrict our expansion range to 
$\mu_q \sim 300 ~\rm MeV$ 
above which the diquark physics is expected to become important.
Also the pion condensation and kaon condensation
takes place in NJL model for $\mu_I > 70~ \rm MeV$ and $\mu_S > 240 ~\rm MeV$
respectively. So we restrict our range within $\mu_I< 70 ~\rm MeV$ and 
$\mu_S < 200 ~\rm MeV$ below $T_C$. However, above $T_C$, approximate 
restoration of chiral symmetry implies that the chiral condensates become
almost zero. So above $T_C$,
we have extended the range of $\mu_I$ and $\mu_S$
for the better fit of the coefficients. Near $T_C$ the 
$\chi^2$ (which is same as the least square here) of the fit varies rapidly
with the variation of range of $\mu_X$ over which the fit was done. So near 
$T_C$ we have fitted the pressure for $1~\rm MeV$ gap of temperature and the 
data points are spaced by $0.1 ~\rm MeV$ of chemical potential for all
temperature values. The least-squares of all the fits came out to be $10^{-10}$ or less. 

 Now we study the behavior of the coefficients $c_2$, $c_4$, $c_6$ and $c_8$
for three sets of chemical potentials for model A and model B. In
figure (\ref{qns}) we show the variation of $c_2$, $c_4$, $c_6$ and $c_8$ with
$T/T_C$ for $\mu_X=\mu_q$ for both models and lattice data.
It can be seen that QNS ($c_2^q$) shows 
an order parameter like behavior. At low 
temperature there are small differences between model A and model B
and model A is much closer to the lattice data. At
high temperature $c_2^q$ for model A reaches almost 98\% of 
its ideal gas value whereas for model B it reaches almost 99\%
of its ideal gas value. However lattice data for $N_\tau=6$ at high 
temperature reaches almost the Stefan-Boltzmann (SB) limit. 
The fourth order derivative $c_4^q$ can be thought of as the susceptibility
of $c_2^q$. The figure shows a peak near $T_C$. Near $T_C$
the model B shows much higher peak than the
model A and the peak of the eight-quark interaction is closer to the 
lattice data. At higher temperature both cases
match very well with the lattice data. But near
and above $2T_C$ lattice value converges to the SB limit, however
both cases of PNJL model are slightly away from the SB limit. 
Note that both $c_2^q$ and $c_4^q$ have only fermionic contribution
in the SB limit. Since, the coupling strength is large enough for $T<2.5T_C$,
a sufficient amount of interaction is present in the system. So, it is 
expected that $c_4^q$ will not converge exactly to the SB limit 
within $T<2.5T_C$. 
The higher order coefficients $c_6^q$ and $c_8^q$ show interesting 
behavior near $T_C$. Although at very low and high temperatures both of them
converge to zero. Near $T_C$, $c_6^q$ shows sharp peaks for both cases.
However for model B the peak is much sharper. 
Similar behavior can be observed for $c_8^q$, which shows more peaks near
$T_C$.
The reason behind the peaks near the transition temperature may
be due to the increase in fluctuation near $T_C$. The sharper peaks of 
model B is probably due to the introduction of enhanced 
repulsive interaction through eight-quark term. This was also reflected 
through the increase of scaled pressure in case of model B Lagrangian
over the model A \cite{deb2}. The number of peaks 
increases near $T_C$ for higher order coefficients.  
\begin{figure}[t]
\centering
\includegraphics[scale=0.9]{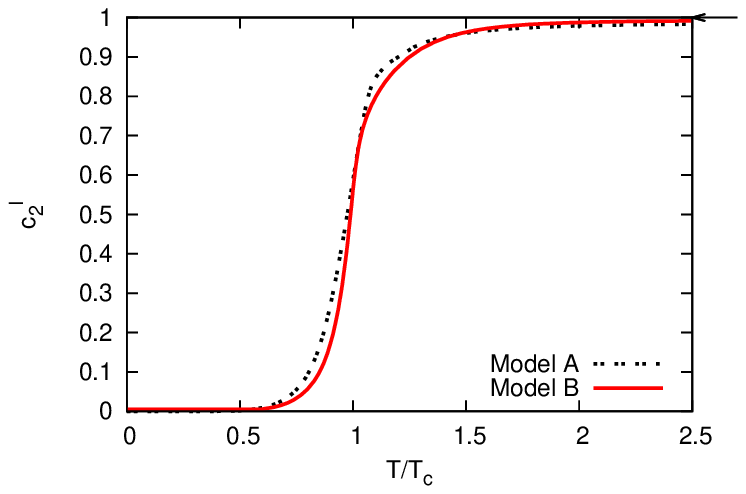}
\includegraphics[scale=0.9]{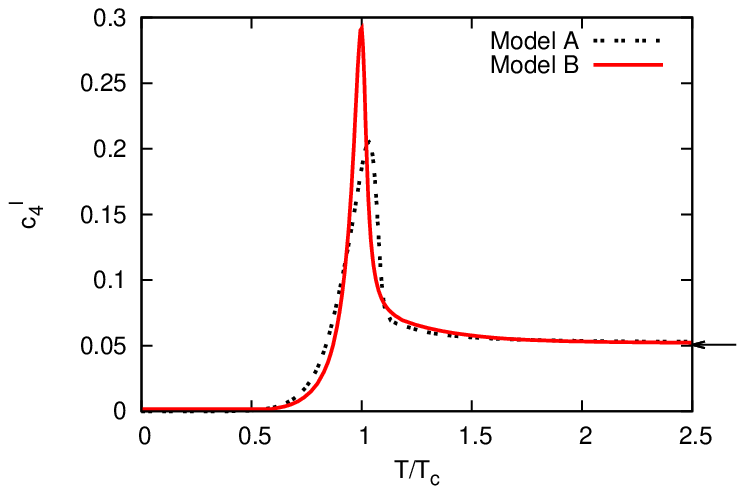}
\includegraphics[scale=0.9]{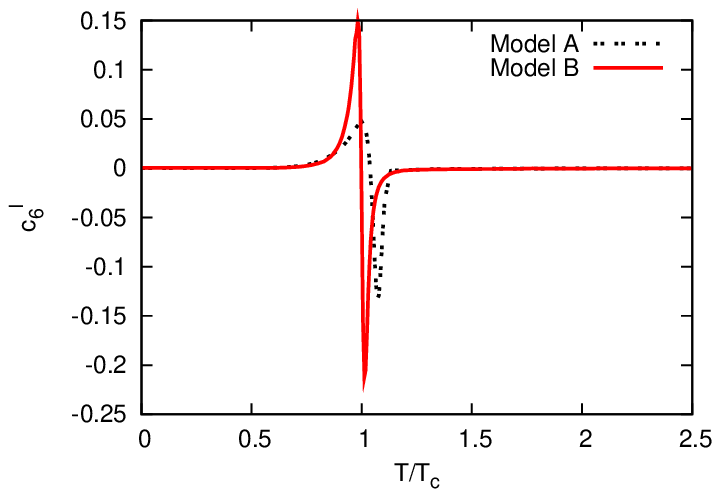}
\includegraphics[scale=0.9]{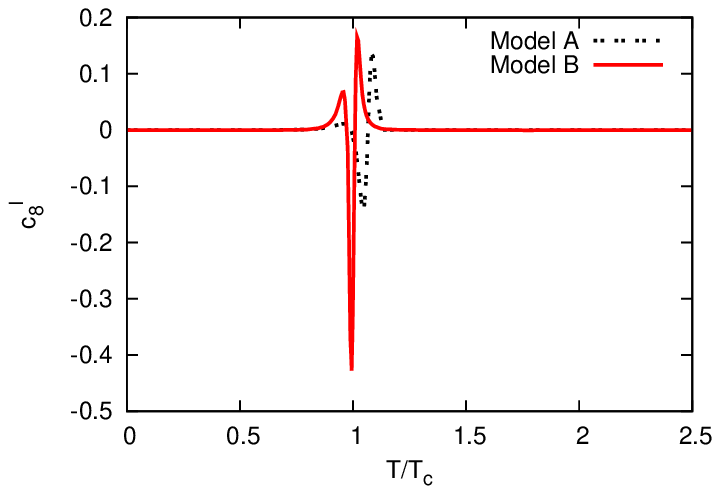}
\caption{(color online). Variation of $c_2$, $c_4$, $c_6$ and $c_8$ 
with $T/T_C$, for $\mu_X=\mu_I$ for models A and B.
The arrows on the right show the corresponding SB limits. The lattice 
data are taken from Ref. \cite{cheng}.}
\label {ins}
\end{figure}
\begin{figure}[t]
\vskip 0.2in
\centering
\includegraphics[scale=0.9]{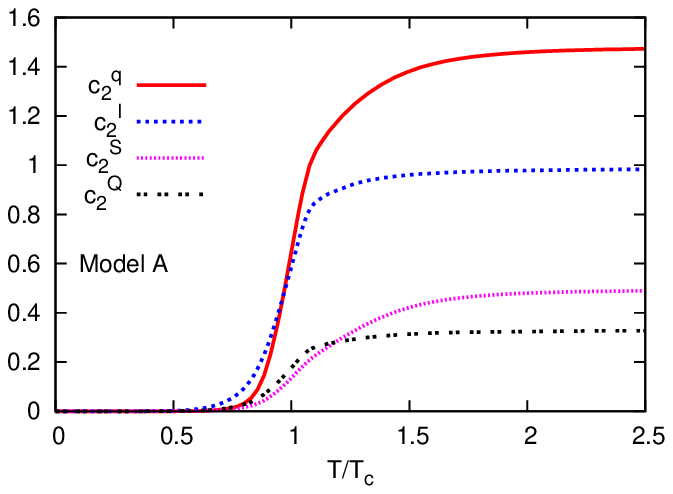}
\includegraphics[scale=0.9]{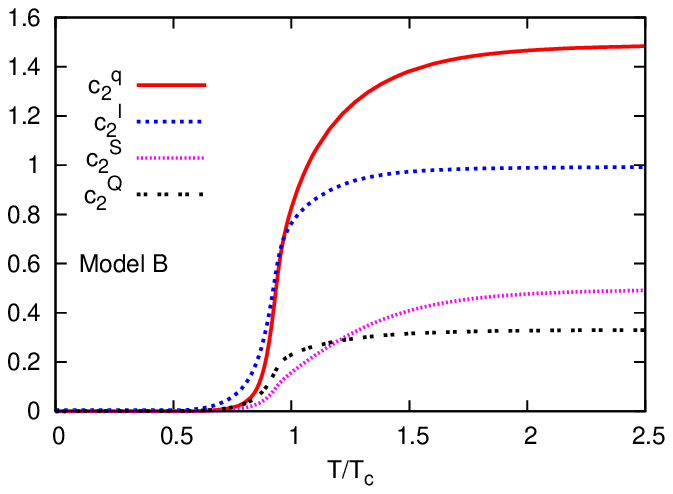}
\includegraphics[scale=0.9]{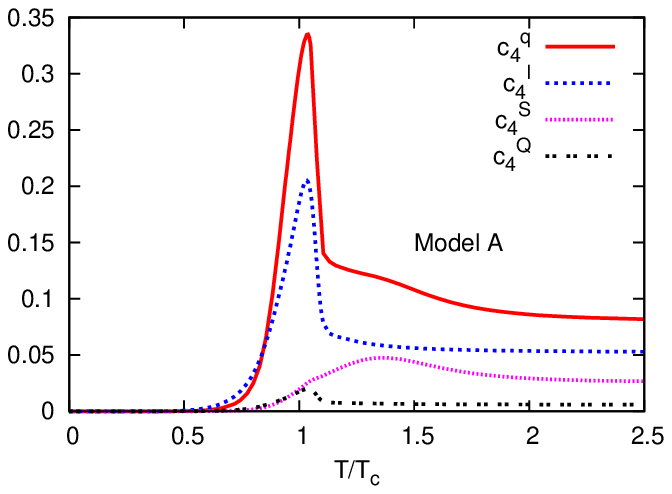}
\includegraphics[scale=0.9]{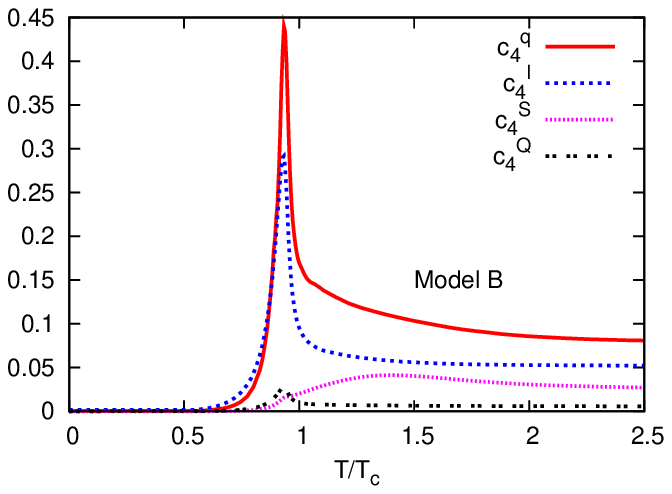}
\includegraphics[scale=0.9]{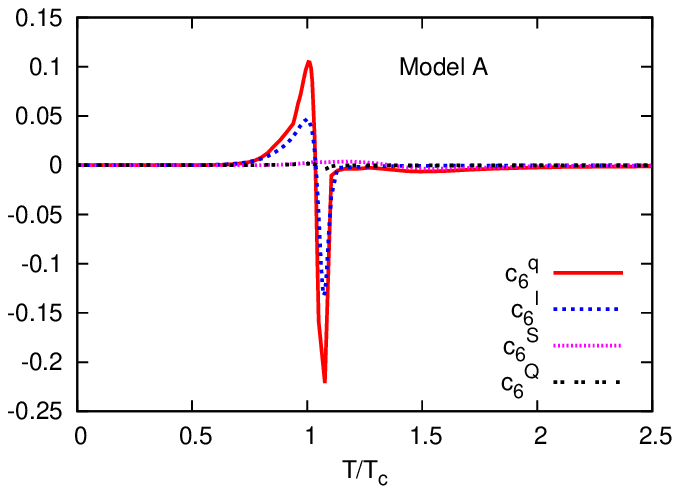}
\includegraphics[scale=0.9]{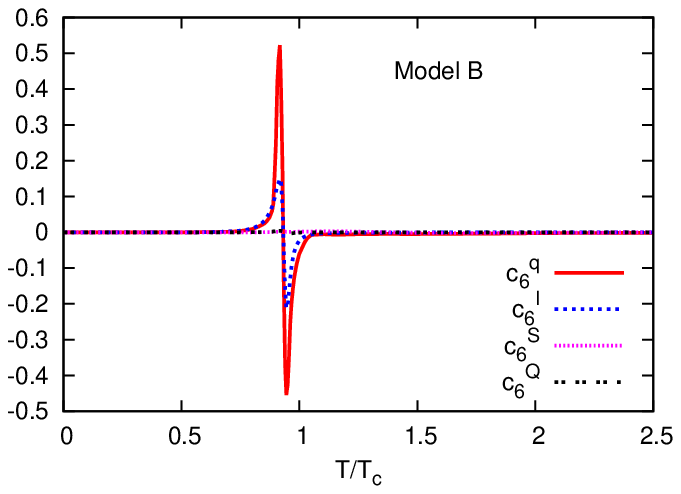}
\includegraphics[scale=0.9]{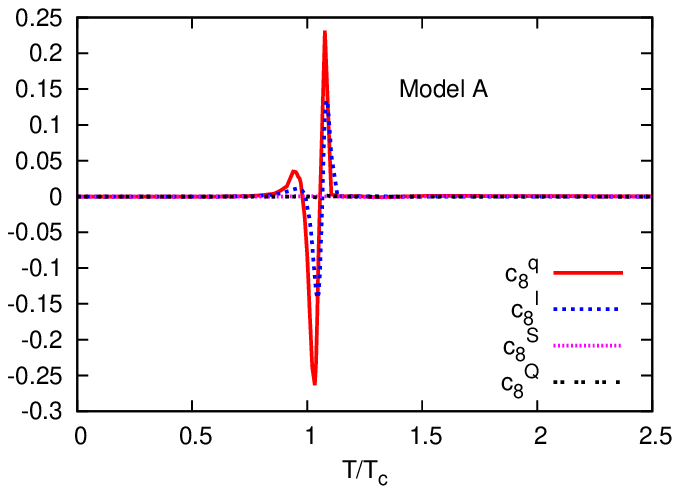}
\includegraphics[scale=0.9]{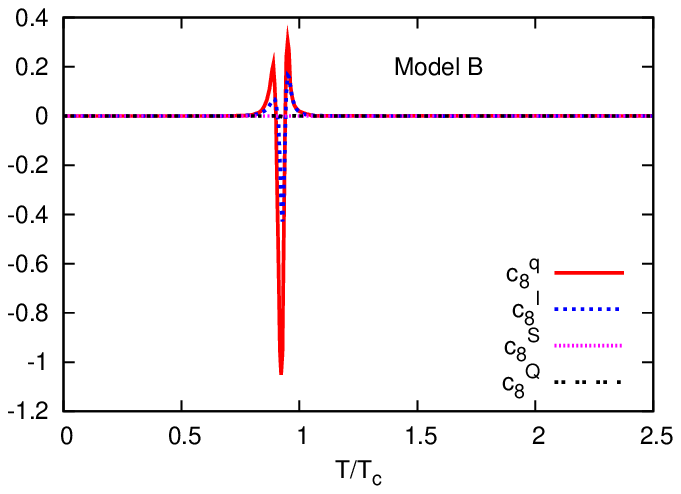}
\caption{(color online). Comparative study of $c_2^X$, $c_4^X$, $c_6^X$ 
and $c_8^X$ for models A and B, where X=q, Q, I or S.}
\label {6qwhole}
\end{figure}
In figure (\ref{charge}) the variation of susceptibility and the higher order 
coefficients for the charge chemical potential is shown. The nature of all the 
coefficients are same as the quark chemical potential. At high temperature the
fluctuation $c_2^Q$ for model B is closer to the SB limit 
compared to the model A. However lattice data is slightly 
above the SB limit for $c_2^Q$. For the case of $c_4^Q$, our data 
(for both cases) show a better 
convergence towards SB limit, unlike $c_4^q$. 
At low temperature the behavior of model A is closer to the
lattice data compared to the model B. The quartic fluctuations
show a peak near $T_C$. The peak for model B is sharper
than model A and the plot for model B
matches well with the lattice result.  The higher order coefficients show 
similar behavior as the quark chemical potential case.
 
  The figure (\ref{sns}) shows the variation of susceptibility and the higher
order coefficients for the strangeness chemical potential with $T/T_C$. The
$c_2^S$ for both the models are slightly different from 
the lattice data. Both the plots are almost 98\%
of the SB limit at high temperature, however the lattice data coincides
with the SB limit. The $c_4^S$ has a similar behavior as the $c_4^q$.
However the peak is not near $T_C$ in this case. Near $T_C$ we 
can see a small bump for both type of Lagrangian, 
but the peak in both case is at higher temperature. This is due to 
the fact that during the chiral crossover the strange quark (the only element 
which carries strangeness) is sufficiently heavy and the corresponding 
condensate $\sigma_s$ melts at much higher temperature than $T_C$.
The maxima of $d\sigma_s /dT$ and $c_4^S$ coincide at the temperature where
peaks are observed. This behavior is quite consistent with the lattice results
which also indicates two peaks. However in case of lattice, the peak at $T_C$ is
higher than the second peak.  So, one can not really pin down the cause of the 
double peak structure. It may be a model artifact. At high temperature both 
the curves are above the SB limit. 
But the lattice data is slightly below the SB limit. 
The higher order derivatives also show a sharp peak at the transition
temperature followed by a broader peak at higher temperature for both model A and Model B. 

For the sake of completeness we have also plotted different moments of pressure 
for isospin chemical potential $\mu_I$ in fig. (\ref{ins}). 
It is clearly seen that, $c_2^I$ has also an order parameter like behavior 
like all other susceptibilities both models. 
At high temperature, the plot for model B
reaches almost 99.5\% of the SB limit, whereas the plot for model A
reaches close to 98\% . For $c_4^I$, the plot for the model B
shows higher peak than the case of model A at $T_C$ and both 
of them converges very well 
towards the SB limit at high temperature. The $c_6^I$ and $c_8^I$ shows 
very rapidly fluctuating peak structure near $T_C$ and goes to zero in both
high and low temperature regime.
\begin{figure}[h]
\centering
\includegraphics[scale=0.9]{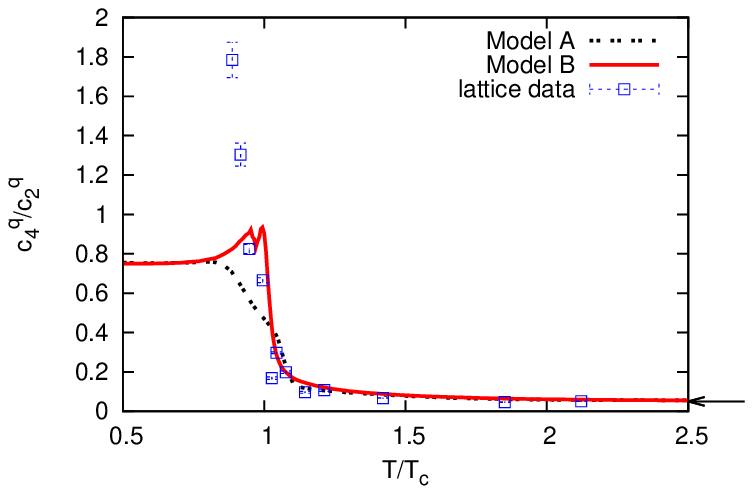}
\includegraphics[scale=0.9]{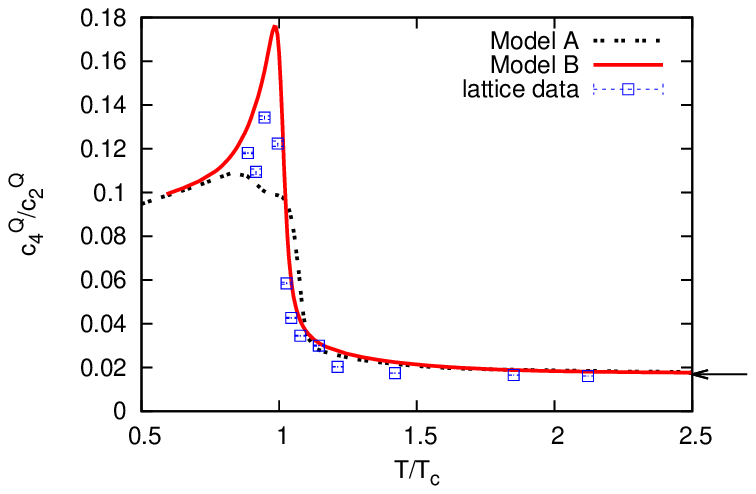}
\includegraphics[scale=0.9]{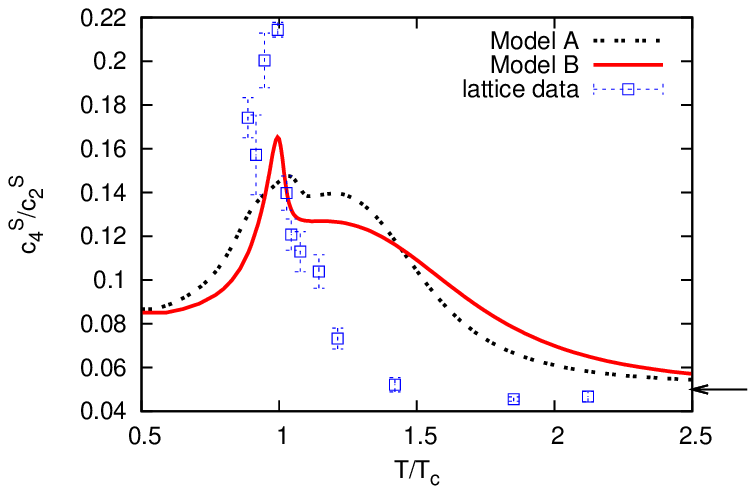}
\caption{(color online). Variation of  $c_4/c_2$ with $T/T_C$, 
for $\mu_q$, $\mu_Q$ and $\mu_S$ for models A and B.
The arrows on the right show the corresponding SB limits. The lattice 
data are taken from Ref. \cite{cheng}.
The upper left panel corresponds to the quark chemical potential, the upper 
right panel corresponds to the charge chemical potential and the lower panel
corresponds to the strangeness chemical potential  }
\label {c4c2_kurt}
\end{figure}
\par
In figure (\ref{6qwhole}) we show a comparative study 
for $c_2$, $c_4$, $c_6$ and $c_8$
with $T/T_C$ for $\mu_q$, $\mu_I$, $\mu_S$ and $\mu_Q$ for both type of 
Lagrangian. In all cases we can see that the quadratic fluctuations rise 
rapidly in the transition region where the quartic fluctuations show a peak.
This peak is most pronounced in case of $\mu_q$. The generic
form of this temperature dependence, a smooth crossover for quadratic 
fluctuations and a peak in quartic fluctuations, is in fact expected to occur
in the vicinity of the chiral phase transition of QCD. For all of the 
coefficients the fluctuation in $\mu_q$ direction is strongest followed by the 
isospin direction fluctuations and the strangeness fluctuation. The charge 
fluctuation is the least pronounced. 
The quartic fluctuation for $\mu_S$ shows a peak at much 
higher temperature than the transition temperature. However other three $c_4^q$,
$c_4^Q$ and $c_4^I$
show peaks at $T_C$. For higher order coefficients the strangeness and charge
fluctuations are negligible near $T_C$ compared to the quark number and 
isospin fluctuations.
\par
In fig (\ref{c4c2_kurt}) we have plotted the kurtosis $\it{i.e.}$ the
ratio of $c_4^X/c_2^X$ for both type of potential (where $X=q, Q ~ \rm{or} ~ S$)
and compared with the lattice data.
The plot for $c_4^q/c_2^q$ for model B
shows more fluctuation near $T_C$ than model A. However 
lattice data shows higher fluctuation near $T_C$ than the model study.
At higher temperature both models coincide with the lattice 
data and converges well with the SB limit.
Our results are qualitatively similar with PQM result
\cite{schaefer3}. In case of the ratio $c_4^Q/c_2^Q$
the model B shows more fluctuation than the model A
and as well as the lattice data. The model B shows
almost 99\% convergence with the SB limit at high temperature. 
For the strangeness fluctuation, we can see two peaks 
in $c_4^S/c_2^S$ curve for the both models.
First peak occurs at chiral transition for light flavors and
second peak occurs when chiral transition occurs in strange
sector. At intermediate temperatures PNJL model overestimates the ratio
than LQCD result. This feature is also observed in PQM
model \cite{schaefer3}. In PQM model the ratio approaches
the SB limit at high temperature and near the transition
region sharp peaks appear. In our case
the ratio shows sharper peak near the transition
temperature and a broader peak at a higher temperature than $T_C$.
However lattice data shows much higher fluctuations than the
plots for both models near $T_C$. At high temperature the model A is closer
to the SB limit than the model B.

\vskip 0.1in
{\subsection{Specific heat and the speed of sound}}
\begin{figure}[h]
\centering
\includegraphics[scale=0.9]{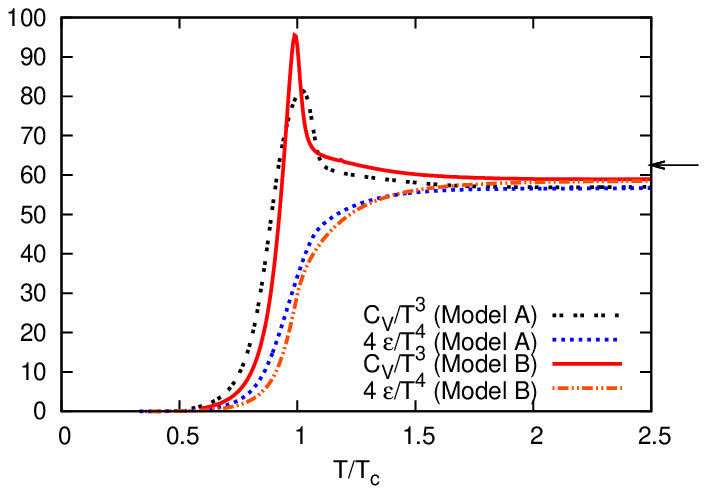}
\includegraphics[scale=0.9]{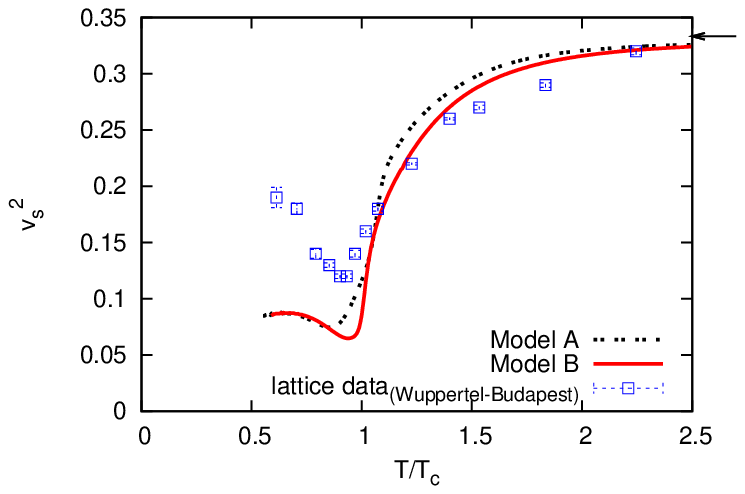}
\includegraphics[scale=0.9]{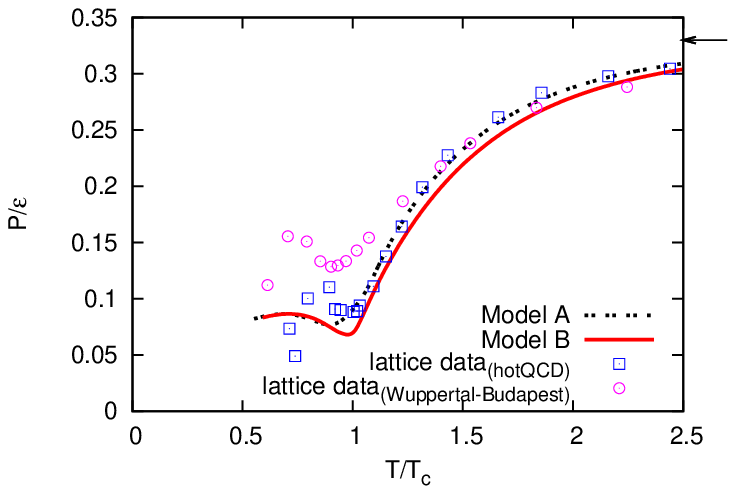}
\caption{(color online). Variation of $C_V/T^3$, $4\epsilon/T^4$,  $v_s^2$ and $p/{\epsilon}$
with $T/T_C$, for models A and B.
The arrows on the right show the corresponding SB limits. The upper left panel shows the plot for $C_V/T^3$ and
$4\epsilon/T^4$ with $T/T_C$,
the upper right panel shows the comparison of $v_s^2$ for models A and B
and the lattice data given by
\cite{fodor10} and the bottom panel shows the comparison of $p/{\epsilon}$
of both models and the lattice data given by \cite{cheng1,fodor10}} 
\label {cvt3}
\end{figure}
We now discuss the thermodynamic quantities like specific heat ($C_V$) and the 
speed of sound ($v_s$). In Fig. (\ref{cvt3}) we have plotted $C_V/T^3$
with $T/T_C$ for both models. From the plot
we can see that $C_V$ grows with increasing temperature and reaches a peak at
$T_C$ for both models. However model B
shows a sharper peak at $T_C$ compared to the 
model A. Just above $T_C$ both the plots
decrease sharply for a short range of temperature. 
Thereafter it gradually converges to a value slightly lower than the 
ideal gas value at high temperature. However the convergence towards the
ideal gas value is better in case of model B. For comparison
we have also plotted the values of $4\epsilon/T^4$, at which the specific heat
is expected to coincide for a conformal gas. From the graph we can see that the
specific heat converges very well with the $4\epsilon/T^4$ at high temperature
for both the models. Both the plots of $C_V/T^3$
and $4\epsilon/T^4$ show similar behavior as PQM model \cite{schaefer3}.
In PQM model $C_V$ grows with temperature and shows a sharp peak at $T_C$. 
After the peak a broad bump is found around 1.2$T_C$ and then $C_V$ goes 
gradually to the ideal gas value.
\par
We now consider the speed of sound and $p/{\epsilon}$ for model A and model B
in figure (\ref{cvt3}). We can see that $v_s^2$ is 
slightly below the ideal gas value at temperature $2.5T_C$ for both cases. 
Our result is quite consistent with the lattice data for 2+1 flavor staggered
fermions reported in \cite{szabo}. A similar behavior as ours is observed in PQM
model \cite{schaefer3}. We get the 
minimum of $v_s^2$ just below the $T_C$ similar to lattice data \cite{szabo}
and the softest point of the equation of state is found to be
$(p/\epsilon)_{min}\approx 0.07$ for model A and 
$(p/\epsilon)_{min}\approx 0.06$ for model B.
Model B gives better agreement with lattice data, which has its softest
point of equation of state as $(p/\epsilon)_{min}\approx 0.05$. 
We have compared two sets of lattice data of Ref. \cite{cheng1} and Ref. \cite{fodor10}
with our model study. Our result shows a better agreement with that of Ref.
\cite{cheng1}. The softest point of equation of state of \cite{cheng1}
is at  $(p/\epsilon)_{min}\approx 0.08$, whereas the softest point of Ref. \cite{fodor10}
is at much higher value $\sim 0.13$. In case of PQM model the softest point of
the equation of state is found to be around $\approx 0.04$ \cite{schaefer3} which is closer to 
our value.

\vskip 0.3in
{\section{Discussion}}
 We have studied the various fluctuations and some of the thermodynamic
quantities using two diferent versions of PNJL model to understand the 
properties of the strongly interacting matter.
In fact it is expected that the susceptibilities and the higher order
fluctuations might provide the direct evidence of the order of the QCD 
phase transition. A pronounced peak in the susceptibility can depict the 
crossover transition and the sharp diverging behavior would indicate the 
existence of a phase transition. 

   We have also obtained the susceptibilities and the higher order 
derivatives by the Taylor expansion of pressure for two kinds of
PNJL model near $\mu_X=0$, where $X=q, I$ or $Q$ and $S$. In all cases the 
second derivative of pressure which is known as the susceptibility, 
show a steep rise near the transition region, which indicates near the 
transition region the fluctuation increases. However at higher temperature
$c_2^X$ almost saturates and almost converges to the ideal gas value. 
This result is quite consistent with the lattice data. The higher
order fluctuation $c_4$ shows a peak near $T_C$ for both models 
and the result matches with the lattice data.
The finite height of the peak confirms 
the crossover nature of transition at $\mu=0$. Both $c_6$ and $c_8$ show rapid
variation around $T_C$ for all cases.

  We have also calculated the specific heat, speed of sound for both 
kinds of PNJL model. The plot for $C_V/T^3$, after showing a peak 
at $T_C$, converges very well to $4\epsilon/T^4$ curve at high $T$.
At high temperature $v_s^2$ almost
reaches its ideal gas value $1/3$ and the softest point of the equation of 
state has a better agreement with lattice result for model B where eight-quark interaction
is taken into account.

 In our formalism, we do not include pion condensate, kaon condensate and 
diquark condensate which may play an important role for higher values of 
chemical potential. Inclusion of those degrees of freedom may improve our result.
Involved though, such studies are in progress.

\vskip 0.2 in
{\section{Acknowledgement}}
P.D. and A.L. would like to thank CSIR for financial support.
A.B. thanks CSIR and UGC (UPE and DRS) for support. The authors 
thank Saumen Datta for some useful discussions.

\end{document}